\documentclass{article}
\usepackage{spconf,amsmath,graphicx}
\usepackage{comment}
\usepackage{booktabs}
\usepackage{amssymb}
\usepackage{multirow}
\usepackage{newtxtext,newtxmath}

\usepackage{hyperref}
\usepackage{cite}
\hypersetup{
    colorlinks = false,
}

\title{Extending Multilingual Speech Synthesis to 100+ Languages without Transcribed Data}
\name{
\textit{
Takaaki~Saeki$^{\, 1,2}$\thanks{$^{*}$This work was carried out as an intern at Google, NY, USA.}~~
Gary~Wang$^{\, 1}$~~
Nobuyuki~Morioka$^{\, 3}$~~
Isaac~Elias$^{\, 4}$~~
Kyle~Kastner$^{\, 1}$~~
Fadi Biadsy$^{\, 1}$~~
}
\\[1mm]
\textit{
Andrew~Rosenberg$^{\, 1}$~~
Bhuvana~Ramabhadran$^{\, 1}$~~
Heiga~Zen$^{\, 3}$~~
Fran\c{c}oise~Beaufays$^{\, 1}$~~
Hadar~Shemtov$^{\, 5}$
}
}
\address{$^1\,$Google, USA \quad $^2\,$The University of Tokyo, Japan \quad $^3\,$Google DeepMind, Japan \\
$^4\,$Google, Israel \quad $^5\,$Google DeepMind, USA
\\[1mm]
\texttt{\small 
takaaki\_saeki@ipc.i.u-tokyo.ac.jp
}
\\
\texttt{\small
\{wgary,~nmorioka,~isaace,~kkastner,~rosenberg,~bhuv,~heigazen,~fsb,~hadar\}@google.com
}
}
\begin{document}
\ninept
\setlength{\abovedisplayskip}{6pt}
\setlength{\belowdisplayskip}{6pt}
\maketitle
\begin{abstract}

Collecting high-quality studio recordings of audio is challenging, which limits the language coverage of text-to-speech (TTS) systems. This paper proposes a framework for scaling a multilingual TTS model to 100+ languages using found data without supervision. The proposed framework combines speech-text encoder pretraining with unsupervised training using untranscribed speech and unspoken text data sources, thereby leveraging massively multilingual joint speech and text representation learning. 
Without any transcribed speech in a new language, this TTS model can generate intelligible speech in $>$30 unseen languages (CER difference of $<$10\% to ground truth).
With just 15 minutes of transcribed, found data, we can reduce the intelligibility difference to 1\% or less from the ground-truth, and achieve naturalness scores that match the ground-truth in several languages.
\end{abstract}

\section{Introduction}\label{sec:intro}
While recent neural text-to-speech (TTS) systems have achieved remarkable performance for resource-rich languages, typical TTS systems cover only a small fraction of the world's nearly 6,000 languages.
TTS systems typically require high-quality transcribed audio data for training, where the audio data consists of studio recordings of phonetically-balanced read speech.
This limits the collection of training data in sufficient amounts to develop TTS systems for low-resource languages.
To address the data collection issue, previous studies have looked at alternative data sources.
Some studies have focused on using unpaired speech-text data~\cite{Ren2019AlmostUT,Chung2019SemisupervisedTF,zhang2020unsupervised,kharitonov2023speak} to reduce the amount of paired data, while others have developed unsupervised TTS methods without paired data by using unsupervised speech recognition~\cite{ni22_interspeech} or a cross-lingual language model~\cite{saeki2023learning}.

We propose a joint speech-text representation learning framework for TTS language expansion. Leveraging a pretrained self-supervised multilingual speech foundation model to define a joint speech-text feature space, we apply both supervised and unsupervised losses to extend to new languages. Pseudo labeling of speech supports learning without manual transcription, while self-supervised text losses mediated through a speech-defined feature space allows for learning from text. The foundation model defined feature space allows us to use a single pretrained audio decoder across all languages. 
Language and speaker ids enable control and cross-speaker and cross-lingual knowledge transfer.
This design enables the framework to flexibly train on \textit{found data}, expanding language coverage of TTS to 100+ languages.

We define \textit{found data} here in contrast to data that has been recorded or curated specifically for training TTS models. 
They include multilingual sources of: speech-text paired data, untranscribed speech data, and unspoken text data. Found speech data includes varied recording conditions, linguistic inconsistencies, imprecise pronunciation, and the presence of disfluencies. This is the kind of data typically used to train ASR and self-supervised (SSL) speech models. With the reduced  training data requirements, \textit{found data} sources have the potential to increase the number of languages covered by TTS to include low resource languages.
Our evaluation results show that the proposed TTS model can generate intelligible speech (CER difference of $<$10\% to ground truth) in more than 30 languages using untranscribed \textit{found data}\footnote{Audio samples are available in \scriptsize{ \url{https://google.github.io/tacotron/publications/extending\_tts}}}.
Our main contributions are:

\begin{itemize} \leftskip -5.5mm \itemsep -0.5mm
    \item A novel TTS framework using unsupervised joint speech-text learning, confirmed to be effective in zero and minimally supervised settings.
    \item Scaling TTS to 100+ languages, spanning multiple language families and writing systems.
    %\item The proposed unified speech model is capable of simultaneously performimg speech synthesis and recognition in a single system. 
    \item By leveraging found data, the proposed TTS model can use SSL speech representations as the target features, thereby enabling the use of an independently trained speech decoder. %br this line was to emphasize use of generic usm encoder iirc
    \item The proposed model, trained on public corpora widely used for multilingual ASR, shows intelligibility differences of 1\% or less when compared to ground-truth data in 30 languages, using around 15 minutes of transcribed found data in the target language.
\end{itemize}

\section{Related Work}\label{sec:related}

Many studies have focused on SSL speech pretraining~\cite{Baevski2020wav2vec2A,Chung2021w2vBERTCC,chiu2022self} and speech-text joint pretraining~\cite{Bapna22mslam,Chen2022MAESTROMS} to improve downstream tasks. The joint speech-text representation has been shown to be effective in developing ASR models without supervised resources~\cite{maestro_u}. Our approach also uses the joint representation for unsupervised TTS. Recent work has employed speech representations from vector-quantized autoencoders~\cite{liu2022delightfultts,chen2023vector} or wav2vec~2.0\cite{siuzdak2022wavthruvec} as the target features of neural TTS models. This approach has been shown to be effective in training a TTS model on a real-world spontaneous speech corpus~\cite{chen2023vector} or in scaling up training data for zero-shot TTS~\cite{wang2023neural}. Our TTS model also exploits multilingual ASR corpora~\cite{pratap20mls,gales14babel,wang21voxpopuli,Conneau2022FLEURSFL} by using BEST-RQ~\cite{chiu2022self}-based target features.

While early work on multilingual TTS~\cite{Zen_SLF_TASLP,li2016multi}  focused on a few resource-rich languages, recent efforts have also focused on expanding multilingual TTS language coverage.
Some have facilitated knowledge transfer from resource-rich to low-resource languages using UTF-8 bytes~\cite{He2021MultilingualBM,saeki2022virtuoso} or articulatory features~\cite{lux2022language}.
In~\cite{saeki2022virtuoso}, the authors propose a multilingual TTS model trained on paired ASR data using joint speech-text semi-supervised learning. Scaling TTS to many
%1000
languages using paired data was proposed in ~\cite{pratap2023scaling} where {\it monolingual, supervised} TTS for each new language was built.  
In this work, we extend the language coverage of a multilingual TTS model through {\it unsupervised learning}.
Previous studies have also used untranscribed speech and unspoken text data to build TTS models with limited paired data~\cite{Ren2019AlmostUT,kharitonov2023speak} or no paired data at all~\cite{ni22_interspeech,saeki2023learning}.
While these studies are limited to evaluation on a few languages, we show that our framework with a {\it single multilingual model} scales to 100+ languages. % br I really dont like 1000 and 100 in the same para
% kk any reason not to swap 1000 languages to "many" above? especially given the tts systems were monolingual... that paper title has 1000 in it but not sure we have to re-list the number here
%ar i dont hate it.

\vspace{-2mm}
\section{Proposed framework}\label{sec:method}
\vspace{-2mm}

\begin{figure}
    \centering
    \includegraphics[width=0.90\linewidth, clip]{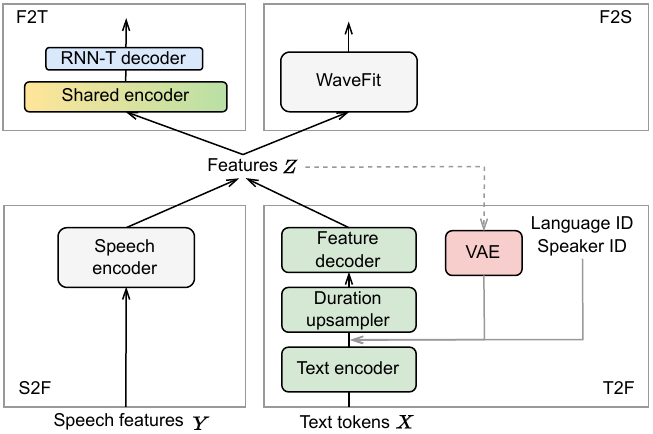}
    \caption{Supervised learning with paired speech-text data.}
    \label{fig:supervised}
    \vspace{-3mm}
\end{figure}

%%%%%%
The proposed multilingual joint speech-text model, as shown in Fig.~\ref{fig:supervised}, includes four main components: speech-to-feature (S2F), feature-to-text (F2T), text-to-feature (T2F), and feature-to-speech (F2S).  
The S2F component and F2T components together form a Conformer~\cite{gulati2020conformer} RNN-T ASR model; the T2F and F2S blocks form a TTS model and are used for inference. 
The S2F components includes the first 6 blocks of a Conformer encoder \cite{gulati2020conformer}. The F2T component contains the remaining 18 conformer blocks, and an RNN-T decoder to predict UTF-8 byte tokens. This structure allows us to leverage a SSL pretrained foundation model~\cite{zhang2023google} to initialize the conformer blocks in both the S2F and F2T components.
The pretrained S2F block is kept frozen after initialization.
This defines a fixed feature space to constrain the remaining components for knowledge transfer across language and modality.
The F2S component is a WaveFit vocoder~\cite{koizumi2023wavefit} that has been trained based on studio-quality American English speech.
This is another foundation component that is reused across all experiments.  
The F2T block is essential for unsupervised training.
Similar to Speech chain~\cite{tjandra17chain}, we use the S2F-F2T path to generate pseudo-labels for unsupervised speech.
For unsupervised text training, we use the T2F-F2T path, mediated by the shared speech-text feature space $Z$.

The T2F component consists of a text encoder, a duration upsampler, a feature decoder and a variational autoencoder (VAE), components that have been used in non-autoregressive TTS modeling \cite{elias2021parallel, elias21_ptaco2}.
As in ~\cite{saeki2022virtuoso,maestro_u}, we use \textit{Bytes} as text tokens using UTF-8 bytes with a vocabulary size of 256.
Bytes facilitate cross-lingual sharing of grapheme representations, while the smaller codebook substantially eliminates the impact of unseen graphemes.
Speaker and language embeddings are additional inputs to the T2F component that are appended to the output of the text encoder prior to duration upsampling.
We use out-of-vocabulary (OOV) IDs for unknown speakers and languages to facilitate transfer and robustness.
This is similar to an architecture with speech, text and shared encoders first introduced for ASR\cite{Chen2022MAESTROMS} and later employed for TTS\cite{saeki2022virtuoso}, here, we use an internal layer of an SSL-trained foundation model to define the joint speech-text feature space.

\vspace{-2mm}
\subsection{Training Objective}\label{sec:method-supervised}
\vspace{-2mm}
The proposed joint speech-text model can be trained on paired (transcribed) speech, untranscribed speech and unspoken text. %The 

During supervised training, an RNN-T decoder generates an alignment between $Z$ and $X$ to upsample $X$ and train the duration predictor~\cite{saeki2022virtuoso}.
Let $\mathcal{L}_{\mathrm{feature}}$ denote the feature loss defined by taking the iterative $L_1$ loss~\cite{elias2021parallel} between the target and predicted features, $Z$ and $\hat{Z}$ respectively.
Let $\mathcal{L}_{\mathrm{rnn-t}}$ and $\mathcal{L}_{\mathrm{dur}}$ denote the RNN-T loss and duration loss, respectively.
We feed $Z$ to the posterior encoder of the VAE, and the KL loss term $\mathcal{L}_{\mathrm{kl}}$ is added to the training objectives.
The training objective for the supervised learning $\mathcal{L}_{\mathrm{sup}}$ can be written as 
\begin{equation}
\mathcal{L}_{\mathrm{sup}} = \mathcal{L}_{\mathrm{feature}} + \mathcal{L}_{\mathrm{kl}} + \mathcal{L}_{\mathrm{dur}} + \mathcal{L}_{\mathrm{rnn-t}},
\end{equation}
where each loss term includes a weighting term.
Since we use SSL speech representations as the target features,
we initialize the speech- and shared-encoder with BEST-RQ pretraining~\cite{chiu2022self}.

\begin{figure}
    \centering
    \includegraphics[width=0.90\linewidth, clip]{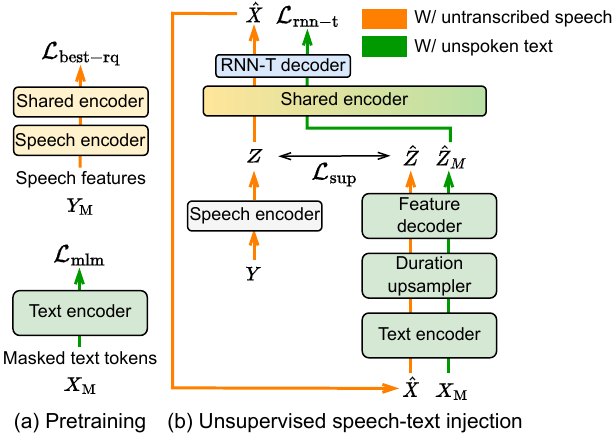}
    \caption{Self-supervised speech-text pretraining and unsupervised speech-text injection using untranscribed speech and unspoken text.}
    \label{fig:unsupervised}
    \vspace{-4mm}
\end{figure}

During inference, a pretrained WaveFit~\cite{koizumi2023wavefit} vocoder directly synthesizes waveform from features $\hat{Z}$, which is predicted by the TTS model component using the VAE latents sampled from the prior.
Note that this WaveFit vocoder is trained to synthesize waveforms from the ground-truth speech-encoder features $Z$.
The vocoder is pre-trained on an American English speech corpus, and is not fine-tuned on the found multilingual data.

In language expansion,
we aim to build TTS on languages which do not have any paired data. To feed the language and speaker IDs during training, we apply classifier-free guidance (CFG)~\cite{ho2022classifier}.
In this CFG technique, the speaker and language embedding corresponding to the OOV IDs can be interpreted as a general speaker and language representation learned from the training data.
This method can reduce biases of particular combinations of language and speaker, a desirable property for multilingual TTS.

\subsection{Speech-text encoder pretraining}\label{sec:method-pretraining}
We use successful pretraining frameworks that facilitate cross-lingual transfer~\cite{conneau2020unsupervised} and language model pretraining~\cite{wu2019beto} to train TTS models with unsupervised data.
As shown in Fig.~\ref{fig:unsupervised}(a), we pretrain the S2F, the Conformer encoder in F2T, and the text encoder in T2F on multilingual unpaired datasets;
\textit{BEST-RQ pretraining}: the Conformer block in S2F and F2T are trained using BEST-RQ training objectives~\cite{chiu2022self}, denoted $\mathcal{L}_{\mathrm{best-rq}}$;
\textit{Text MLM pretraining}: The text encoder in T2F are trained using multilingual BERT pretraining objectives~\cite{devlin2019bert}, denoted $\mathcal{L}_{\mathrm{mlm}}$, where input text tokens are masked with span of 20 tokens, at 15 percent, similar to \cite{song2019mass}.
%As presented in \S~\ref{sec:method-curriculum}, 
We use these pretrained encoders to initialize the TTS model.

\subsection{Joint training with unsupervised speech-text data}\label{sec:method-injection}

Three types of multilingual training data are used to achieve zero supervision on new languages; speech-text paired data, \textit{untranscribed speech data} (no corresponding text), and \textit{unspoken text data} (no corresponding speech). We use joint speech-text unsupervised learning methods to take advantage of these sources as shown in Fig.~\ref{fig:unsupervised}(b).

\noindent \textbf{Unspoken text.} %While we perform pretraining with unspoken text data as described in \S~\ref{sec:method-pretraining}, 
We use text-only unsupervised learning criteria for joint training, which is similar to text-only training from previous work~\cite{Chen2022MAESTROMS,maestro_u}. \textit{Aligned text MLM}: Let $X_{\mathrm{M}}$ and $Z_{\mathrm{M}}$ denote the input masked text tokens and the generated SSL features, respectively.
%As shown in Fig.~\ref{fig:unsupervised}(b), 
Note that the input masked text tokens are upsampled using the trained duration predictor.
$\mathcal{L}_{\mathrm{rnn-t}}$ is computed through the shared encoder and RNN-T decoder. $\mathcal{L}_{\mathrm{rnn-t}}$ is back-propagated to the TTS part to train the TTS model using unspoken text data.
Since the target speech is not available, we use the latents sampled from the prior and the unknown speaker IDs for the TTS model.

\noindent \textbf{Untranscribed speech.} For untranscribed speech training, we use the following unsupervised training criteria;%\textit{Pseudo-labelling} and \textit{Speech MLM}. 

\noindent \textit{Pseudo-labeling}: %As shown in Fig.~\ref{fig:unsupervised}(a), 
Hypothesis $\hat{X}$ is decoded from untranscribed speech representations $Z$ from S2F and F2T.
$\hat{X}$ is then fed into T2F to train its components using untranscribed speech. We use $\mathcal{L}_{\mathrm{sup}}$ (Section ~\ref{fig:supervised}) to train the T2F module. We apply stop gradient to the output of the RNN-T decoder.
Note that these pseudo-labels could also be generated by an external ASR model. %, which we leave for future investigation.  
We evaluate the contribution of this unsupervised training scheme in Section ~\ref{sec:eval-ablation}.
The relationship between the accuracy of the pseudo-labels and the resulting TTS performance remains a topic for future work.

\subsection{Curriculum training procedures}\label{sec:method-curriculum}
We use a curriculum learning procedure in three stages. The first stage consists of pretraining of the speech and shared encoders as described in Section~\ref{sec:method-pretraining}. The second stage consists of freezing the speech encoder, and training only the shared encoder and  the RNN-T decoder using $\mathcal{L}_{\mathrm{rnn-t}}$. This step provides a reliable RNN-T decoder for the third stage. In the third stage, we perform joint training using the supervised learning described in Section~\ref{sec:method-supervised} and the unsupervised speech-text injection described in Section~\ref{sec:method-injection}.

% \vspace{-2mm}
\section{Experimental Setting}\label{sec:eval}
% \vspace{-1mm}

\noindent \textbf{Dataset:} There are three types of training data: speech-text paired data, untranscribed speech data, and unspoken text data.
Language labels are annotated for all the data types, and the speaker labels are included in some of the paired data\footnote{VoxPopuli and MLS corpora include speaker annotation,  others do not.}.
We use two language groups, \textit{Group A} and \textit{Group B}, as categorized by FLEURS~\cite{Conneau2022FLEURSFL} and used in previous studies~\cite{maestro_u}.
Group A and B contain 52 and 50 languages respectively, both of which include multiple language families.
Each language in Group A has at least 10 hours of paired data, and it also has unpaired speech and text data. 
We evaluate Group B languages with no paired data, and then evaluate how performance improves with 15 minutes of paired found data.
While typical multilingual TTS approaches focus on supervised data (i.e. \textit{Group A} languages), we aim to develop TTS models for \textit{Group B} without supervision.

For the paired data used in the supervised learning, %(\S~\ref{sec:method-supervised}), 
the subsets of Group A languages from MLS~\cite{pratap20mls}, VoxPopuli~\cite{wang21voxpopuli}, Babel~\cite{gales14babel}, and FLEURS~\cite{Conneau2022FLEURSFL} were used.
The speech- and shared-encoder pretrained foundation model described in \S~\ref{sec:method-pretraining}, was trained on 20M hours of untranscribed YouTube speech audio consisting of 56 languages from both groups.  
%br i introduced foundation model but we should call this out in contributions
Note that this foundation model was pretrained for an ASR task \cite{zhang2023google}, and reused here.
For the feedforward text encoder pretraining and the unsupervised text training, %(\S~\ref{sec:method-injection}),
we used MC4~\cite{xue20mt5} covering 101 languages including Group A and Group B languages.
For the unsupervised speech training,% described in \S~\ref{sec:method-injection}, 
we used untranscribed speech corpora: MLS, VoxPopuli, Babel, and FLEURS, which contained languages from both groups.
For the training of the WaveFit model, %described in %\S~\ref{sec:method-supervised}, 
we used a proprietary high-quality audio-only dataset including American English only.
For the TTS evaluation, we used 300 utterances sampled from the FLEURS test set for each language.

\noindent \textbf{Model Details:} The S2F and F2T Conformer encoders consist of 6 and 18 Conformer layers, respectively, and the hidden dimension was set to 768.
The feedforward text encoder consisted of 12 Conformer layers with the hidden dimension of 768.
The TTS model follows that of~\cite{elias2021parallel, elias21_ptaco2} with a feature decoder consisting of 6 conformer layers.
The training steps for ASR model training and joint training were 200k and 70k steps, respectively with a batch size of 256. For speech pretraining described in %\S~\ref{sec:method-pretraining}, 
the model was trained for 1.1M iterations with a batch size 4,096.
For the text MLM pretraining, %described in \S~\ref{sec:method-pretraining}, 
the model was trained for 2M iterations with the batch size of 1,024. The KL loss term is trained with a schedule from 6k-50k steps as in \cite{elias21_ptaco2}.

\begin{table}[tb]
\centering
\caption{Comparison results. CER and SQuId are averaged over 50 Group B languages, while MOS are averaged over a 4 language subset of Group B. See \S~\ref{sec:results-setting} for each method description.}
\label{tab:average}
\begin{tabular}{l|ccc}
\toprule
                 & \multicolumn{1}{c}{MOS} & CER (\%)      & SQuId \\ \midrule
\textit{Groundtruth}      &        3.673         & 6.55  & 3.64  \\
\textit{Supervised}           &          3.213         & 6.39  & 3.88   \\\midrule
\textit{Zero Baseline}    &             2.476      & 28.28 & 3.84  \\
\textit{Zero Proposed}        &              2.531         & 23.44 & 3.77  \\ \midrule
\textit{15m Baseline} &              2.928         & 11.17 & 3.91  \\
\textit{15m Proposed}     &                3.180     & 7.33  & 3.88  \\ \bottomrule
\end{tabular}
\vspace{-3mm}
\end{table}

\begin{figure}
    \centering
    \includegraphics[width=0.88\linewidth, clip]{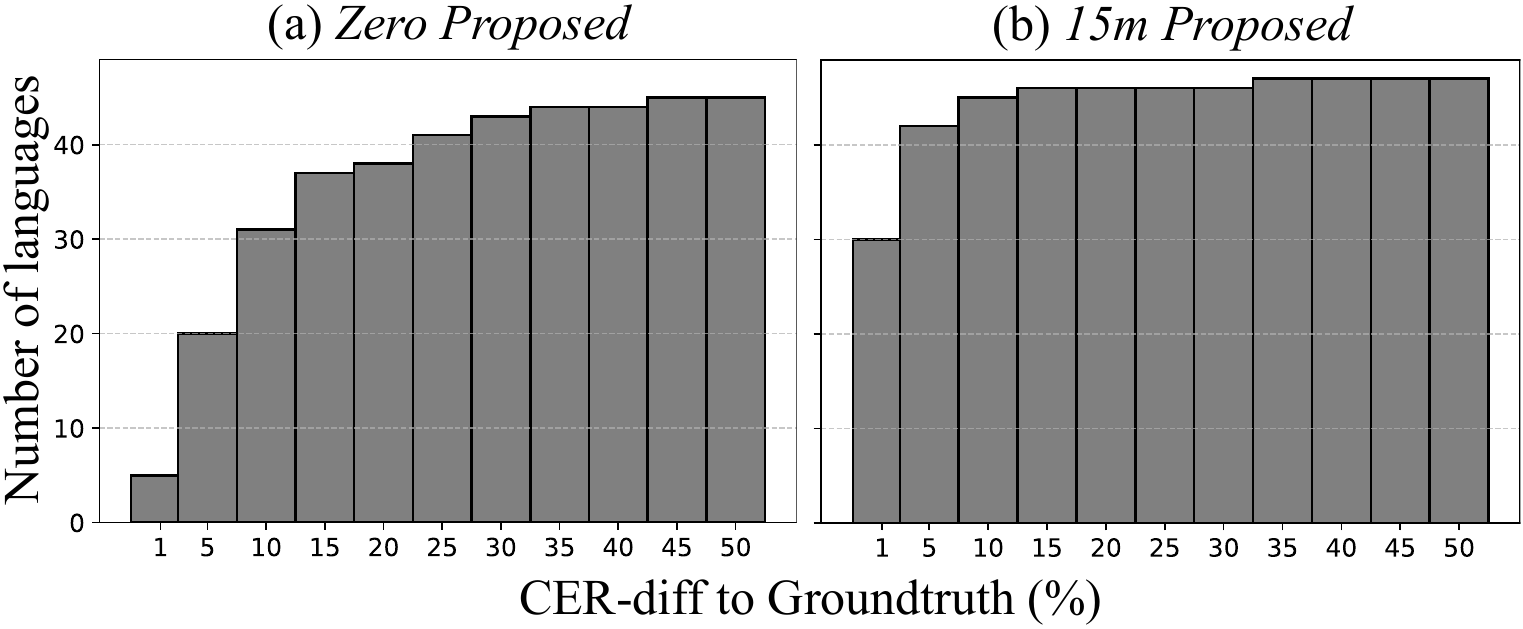}
    \caption{CER difference to groundtruth data, and number of languages.}
    % \vspace{-3mm}
    \label{fig:cer_hist}
\end{figure}

\noindent \textbf{Evaluation metrics:}
For the evaluation of intelligibility, the character error rates (CER) were computed using a pretrained multilingual ASR model~\cite{zhang2023google}.
For the evaluation of naturalness, we conducted subjective evaluations using 5-scale mean opinion score (MOS) in naturalness.
Since it was difficult to have a sufficient number of evaluators for many lower-resource languages, we conducted the MOS tests for four languages: Spanish, Portuguese, %Swedish, 
Arabic, and Tamil, with a total of around 500 raters. %, and Ukrainian. 
As an additional metric for naturalness, we used SQuId~\cite{SQuId}, a pretrained automatic speech quality assessment model.

% \vspace{-2mm}
\section{Results}\label{sec:results-setting}
% \vspace{-1mm}
While our primary goal is zero paired data, we are also interested in how performance scales with \textit{some} available training data.
Therefore, we compared three training data conditions for Group B: \textit{Zero}, \textit{15m} and \textit{Supervised}. \textit{Baseline} model uses bytes as input tokens and Best-RQ pretraining
Condition \textit{Zero}, does not use any paired data from Group B languages.
For Condition \textit{15m}, we extracted about 15 minutes of paired training data from FLEURS for each of Group B languages.
Note that this paired data does not contain speaker labels and consists of noisy audio data with a ground-truth MOS of less than 4.
In \textit{Supervised}, we used all the paired data from MLS, VoxPopuli, Babel and FLEURS for Group B, as in Group A.

\begin{table}[tb]
\centering
\caption{List of languages under different CER-diff conditions.}
\label{tab:language-extension}
\scalebox{0.88}{
\begin{tabular}{ll}
\toprule
\multicolumn{1}{l|}{Condition}                                & Language codes following FLEURS~\cite{Conneau2022FLEURSFL}                \\ \midrule
\multicolumn{2}{l}{\textit{Zero Proposed}}                                                                         \\ \midrule
\multicolumn{1}{l|}{$\text{CER-diff.} <= 1\%$}               & es ff jv ny umb                                \\
\multicolumn{1}{l|}{$1\% < \text{CER-diff.} <= 5\%$} & ast bg fil gu ig lg mr ru sl sn tg tr uz xh zu \\
\multicolumn{1}{l|}{$\text{CER-diff.} > 50\%$}              & he hy ja lo my                                 \\ \midrule
\multicolumn{2}{l}{\textit{15m Proposed}}                                                                      \\ \midrule
\multicolumn{1}{l|}{$1\% < \text{CER-diff.} <= 5\%$} & ar as ca da gu he is kn mr nb ru sd            \\
\multicolumn{1}{l|}{$\text{CER-diff.} > 10\%$}              & cmn ja lo hy                                   \\ \bottomrule
\end{tabular}
}
% \vspace{-2mm}
\end{table}
% \vspace{-2mm}
\subsection{Main results for TTS evaluations}\label{sec:eval-main}
% \vspace{-2mm}
Table~\ref{tab:average} lists the results for MOS, CER and SQuId for the Group B languages.
Note that  CER and SQuId metrics are averages over all 50 Group B languages, while the MOS values are calculated by averaging over four languages.
We used these automated evaluations to identify a candidate for subjective MOS evaluation. The \textit{Proposed} model with text MLM, aligned text MLM, and pseudo-labeling (model (4) in Table ~\ref{tab:ablation}) showed the highest overall intelligibility as measured by CER. Therefore we used this model for MOS evaluations.

In the intelligibility evaluations, we observed that the joint use of the different speech-text unsupervised learning methods reduced the CER by 8.38\% and 9.83\% for the \textit{Zero} and \textit{15m} conditions, respectively.
\textit{15m Proposed} showed a difference of 0.77\% against \textit{Groundtruth} and a difference of 1.03\% against \textit{Supervised}.
This indicates that the proposed method can significantly improve intelligibility by using 15 minutes of transcribed found data.

Also, the best methods showed the SQuId improvement of 0.07 for both \textit{Zero} and \textit{15m} conditions.
It should be noted that, the SQuId score for \textit{Groundtruth} was 3.64 because we used the FLEURS test set for our evaluation, %as mentioned in \S~\ref{sec:eval-setting}.
\textit{15m Proposed}, although inferior to \textit{Supervised}, shows a difference of around 0.04 with \textit{Supervised} and a higher SQuId than the original test data for FLEURS.

\begin{table}[tb]
\centering
\caption{Ablation study at \textit{Zero} and \textit{15m} setting.}
\label{tab:ablation}
\scalebox{0.90}{
\begin{tabular}{clcc}
\toprule
\multicolumn{1}{c|}{\multirow{2}{*}{Index}} & \multicolumn{1}{l|}{\multirow{2}{*}{Method}}    & \textit{Zero}                            & \textit{15m}        \\
\multicolumn{1}{c|}{}                       & \multicolumn{1}{l|}{}                           & \multicolumn{1}{c|}{CER / SQuId}  & CER / SQuId  \\ \midrule
\multicolumn{1}{c|}{(1)}                    & \multicolumn{1}{l|}{\textit{Byte-based Baseline}}                & \multicolumn{1}{c|}{28.28 / 3.84} & 11.17 / \textbf{3.91} \\ \midrule
\multicolumn{4}{l}{\textit{Unspoken text}}                                                                                                   \\ \midrule
\multicolumn{1}{c|}{(2)}                    & \multicolumn{1}{l|}{(1) + \textit{Text MLM pretraining}} & \multicolumn{1}{c|}{26.13 / 3.86} & 8.47 / 3.89  \\
\multicolumn{1}{c|}{(3)}                    & \multicolumn{1}{l|}{(2) + \textit{Aligned text MLM}}     & \multicolumn{1}{c|}{27.90 / \textbf{3.87}} & 8.35 / 3.84  \\ \midrule
\multicolumn{4}{l}{\textit{Untranscribed speech}}                                                                                            \\ \midrule
\multicolumn{1}{c|}{(4)}                    & \multicolumn{1}{l|}{(3) + \textit{pseudo labeling}}      & \multicolumn{1}{c|}{\textbf{23.44} / 3.77} & \textbf{7.33} / 3.88  \\ \bottomrule
\end{tabular}
}
\vspace{-2mm}
\end{table}
%\vspace{-2mm}
\subsection{Analysis for TTS language expansion}\label{sec:eval-num-lang}
%\vspace{-2mm}

For each of \textit{Zero Proposed} and \textit{15m Proposed}, we count the number of languages for which the difference in CER from the ground-truth (referred to as ``CER-diff'') was less than a certain threshold.
Fig.~\ref{fig:cer_hist} shows the histograms.

In the \textit{Zero Proposed}, five languages had a CER-diff of 1\% or less, 20 languages had a CER-diff of 5\% or less, and 31 languages had a CER-diff of 10\% or less.
Table~\ref{tab:language-extension} lists the number of languages in the different CER-diff ranges.
The 20 languages with a CER-diff of 5\% or less included several language families.
The languages with a CER-diff greater than 50\% included tonal languages (e.g., Lao), and languages with many unwritten diacritics (e.g., Hebrew).

From the \textit{15m Proposed} results, the number of languages with the CER-diff of 1\% or less was 30, with 42 languages with the CER-diff of 5\% or less and 45 languages with the CER-diff of 10\% or less.
Languages with a CER-diff greater than 10\% included tonal languages such as Cantonese.
We found that the use of 15 minutes of transcribed found data can significantly improve the intelligibility of Hebrew and Burmese.

% \vspace{-2mm}
\subsection{Ablation studies}\label{sec:eval-ablation}
%\vspace{-2mm}
We conducted ablation studies for the proposed unsupervised speech-text learning. %described in Sections ~\ref{sec:method-pretraining} and \S~\ref{sec:method-injection}.
Table~\ref{tab:ablation} lists the results.
We found that the model~(4) with feedforward text MLM, aligned MLM and pseudo-labeling had the highest intelligibility, demonstrating the effectiveness of the joint speech-text unsupervised learning.
These ablations show the value of being able to train on unpaired text (models~(2) and (3)) and untranscribed speech (model~(4)).
Intelligibility is improved by both.
Moreover, the ability of the model to benefit from these sources is magnified by a small amount (15m) of transcribed speech in the language.
The CER improvement in the Zero setting is about 17\% relative, but 34\% relative in the 15m setting.

%\vspace{-2mm}
\section{Conclusions}
%\vspace{-2mm}
This study proposed a framework for TTS language expansion using joint speech-text unsupervised learning and a single multilingual TTS model.
We used the proposed framework to develop a TTS model for 100+ languages with zero or minimal supervision of the found data.
Without any transcribed speech in a new language, this TTS model generated intelligible speech in >30 unseen languages (CER difference of <10\% to ground truth).
With just 15 minutes of transcribed, found data we reduced the intelligibility difference of 1\% or less from the ground-truth, and achieved naturalness scores matching the ground-truth in several languages.

% References should be produced using the bibtex program from suitable
% BiBTeX files (here: strings, refs, manuals). The IEEEbib.bst bibliography
% style file from IEEE produces unsorted bibliography list.
% -------------------------------------------------------------------------

\clearpage
\bibliographystyle{IEEEbib}
\footnotesize 
\bibliography{refs}

\end{document}